\documentclass[aps,pre,reprint,amsmath,amssymb,superscriptaddress,floatfix]{revtex4-2}
\usepackage{graphicx,xcolor,hyperref}
\usepackage{braket}
\usepackage[ignoreunlbld,norefs,nocites]{refcheck}
\begin{document}

\title{Good-$\Gamma$ Numbers: A Shell-Optimized Route to the Thermodynamic Limit in Quantum Monte Carlo Calculations of the High-Density Two-Dimensional Electron Gas}
\author{S.\ Azadi}
\email{sam.azadi@manchester.ac.uk}
\affiliation{Department of Physics and Astronomy, University of Manchester, Oxford Road, Manchester M13 9PL, United Kingdom}
\affiliation{Department of Physics, Clarendon Laboratory, University of Oxford, Parks Road, Oxford OX1 3PU, UK}
\affiliation{Center for Advanced Systems Understanding, Untermarkt 20, D-02826 G\"orlitz, Germany}
\author{T.\ D.\ K\"{u}hne}
\affiliation{Center for Advanced Systems Understanding, Untermarkt 20, D-02826 G\"orlitz, Germany}
\affiliation{Helmholtz Zentrum Dresden-Rossendorf, Bautzner Landstra{\ss}e 400, D-01328 Dresden, Germany}
\affiliation{TU Dresden, Institute of Artificial Intelligence, Chair of Computational System Sciences, N\"othnitzer Stra{\ss}e 46 D-01187 Dresden, Germany}
\author{A.\ Principi}
\affiliation{Department of Physics and Astronomy, University of Manchester, Oxford Road, Manchester M13 9PL, United Kingdom}
\author{M.\ S.\ Bahramy}
\affiliation{Department of Physics and Astronomy, University of Manchester, Oxford Road, Manchester M13 9PL, United Kingdom}
\date{\today}

\begin{abstract}

Single-particle finite-size errors remain a central obstacle in quantum Monte Carlo (QMC) simulations of periodic systems, conventionally reduced via twist averaging (TA). Here we introduce Good-$\Gamma$ (GG) particle numbers as system sizes $N$ at which the $\Gamma$-point Hartree--Fock (HF) energy of the high-density two-dimensional uniform electron gas (2D UEG) lies anomalously close to its thermodynamic-limit (TL) value. All results in this work concern the 2D UEG in its high-density Fermi-liquid regime, $r_s=1$--$4$. We systematically investigate the kinetic- and exchange-energy errors, $\Delta KE(N)$ and $\Delta EX(N)$, that underlie this one-body finite-size-error cancellation across four densities and both spin polarizations. We then compare diffusion Monte Carlo (DMC) energies extrapolated to the TL from GG-$N$ sequences against twist-averaged extrapolations at the same densities and polarizations, finding agreement to within a fraction of a millihartree in every case, independent of whether a $1/N$ or an $N^{-5/4}$ extrapolation form is used.
\end{abstract}

\maketitle

\section{Introduction}

Quantum Monte Carlo (QMC) simulations, including variational and diffusion quantum Monte Carlo (VMC and DMC), of periodic systems are usually performed in finite simulation cells subject to periodic boundary conditions. As a result, they suffer from finite-size (FS) errors, which arise from using a finite number of particles and a finite simulation-cell volume to represent the thermodynamic limit \cite{Foulkes01,Holzman2011,Holzmann16,Neil08,Azadi15,Azadi19}. Controlling and correcting these errors is essential for applying QMC methods to model Hamiltonians such as the electron liquid, and to real periodic systems, particularly when high accuracy is required.

FS errors are conventionally separated into two categories. The first consists of independent-particle, or one-body, FS errors, associated with the discretization of momentum states in a finite simulation cell. Analogous errors appear in density-functional theory (DFT) calculations of periodic systems and are reduced there by dense $k$-point sampling of the Brillouin zone (BZ); an alternative, dating back to Baldereschi's mean-value point \cite{Balder73} and the special $k$-point methods of Rajagopal \emph{et al.}\ \cite{Raja94,Raja95}, is to replace the full BZ sum with one or a few carefully chosen points. In QMC, the analogous procedure to dense sampling is twist averaging (TA), in which the energy is averaged over boundary-condition phases (twists) to reduce shell-filling effects \cite{Lin01}. The special-point idea has itself been developed further within QMC: Kwee, Zhang, and Krakauer \cite{Kwee08} used a mean-field proxy to construct an external, post-processing FS correction, while Dagrada \emph{et al.}\ \cite{Dagrada16} introduced the ``exact special twist'' (EST) method, which continuously adjusts a single twist angle so that the mean-field energy at that twist reproduces the thermodynamic limit (TL) value to arbitrary numerical precision, for a fixed $N$. The present work is closely related to this line of research but explores a different complementary lever. Rather than tuning the twist at fixed $N$, we fix the twist at $\Gamma$ and search over $N$ for particle numbers at which the $\Gamma$-point HF energy happens to lie close to its TL value. We refer to such $N$ as Good-$\Gamma$ (GG) particle numbers. As we show below (Sec.~\ref{sec:range_100_500}), selecting GG numbers usefully for thermodynamic-limit extrapolation requires more than ranking candidate $N$ by $|\Delta E^{\rm HF}(N)|$ alone, as the selected $N$ must also be reasonably spread in $1/N$, since the deepest cancellations can otherwise cluster at nearly adjacent $N$, leaving a linear fit with almost no leverage. We take this spacing requirement to be part of a practically useful definition of Good-$\Gamma$ numbers, not an optional refinement.

The second category of FS error consists of many-body contributions, arising from the long-range nature of the Coulomb interaction, the finite extent of the exchange-correlation hole, and the artificial periodicity of the simulation cell. Unlike one-body FS errors, many-body FS errors are not removed by twist averaging and require separate correction schemes or extrapolation to the TL \cite{Chiesa06,Fraser96,Kent99,Williamson97,Raja95,Raja94}.

In this work we restrict our attention to the two-dimensional uniform electron gas (2D UEG) \cite{Azadi24} in its high-density Fermi-liquid regime, $r_s=1$--$4$, which we regard as a controlled, model-system setting for investigating why a HF-level shell-cancellation criterion behaves the way it does, and what it can and cannot be expected to deliver at the correlated level -- not as a proposal for a general-purpose replacement for twist averaging. We demonstrate the GG approach at $r_s=1,2,3,4$, for both unpolarized ($\zeta=0$) and fully spin-polarized ($\zeta=1$) polarization, at both the HF and the correlated diffusion quantum Monte Carlo (DMC) level, and, by comparing HF, $\Gamma$-point DMC, and TA-DMC calculations as functions of system size, determine precisely what the GG criterion does and does not achieve.

\section{Model and computational method}
\label{sec:methods}
We consider the 2D UEG, at both unpolarized ($\zeta=0$) and fully spin-polarized ($\zeta=1$) polarization, characterized by the density parameter $r_s$, using hexagonal simulation cells with periodic boundary conditions. We used standard Slater-Jastrow-Backflow trial wave functions, described in detail in our previous work~\cite{Azadi24,Neil04,Pablo12}. Unlike that previous work~\cite{Azadi24}, we did not include the $p$ and $\pi$ terms in the Jastrow and backflow functions, in order to reduce computational cost. The wave functions were optimized by variance minimization~\cite{Neil05} followed by energy minimization~\cite{Toulouse}. We used a target population of 1920 walkers in our DMC calculations, with a time step of 0.01\,(1/Ha); the same walker population and time step were used at both polarizations, $\zeta=0$ and $\zeta=1$, at each density. Twist-averaged energies were obtained using 100 random twists, shifted from $\Gamma$ and drawn from a uniform distribution over the BZ, for $N\leq200$, and 50 twists for $N>200$. We did not repeat an explicit twist-count convergence test for the present calculations; this choice is based on our previous systematic studies of finite-size and twist-averaging behaviour in the uniform electron gas~\cite{Azadi19,Azadi21,Azadi22,Azadi23,Azadi23II,Azadi24}, where this number of twists was established to be sufficient for the densities and system sizes considered here. All QMC calculations were performed using the \textsc{casino} package~\cite{casino}. 

\section{Hartree-Fock}
\label{sec:hf}

\begin{figure}[t]
    \centering
    \includegraphics[width=1.0\linewidth]{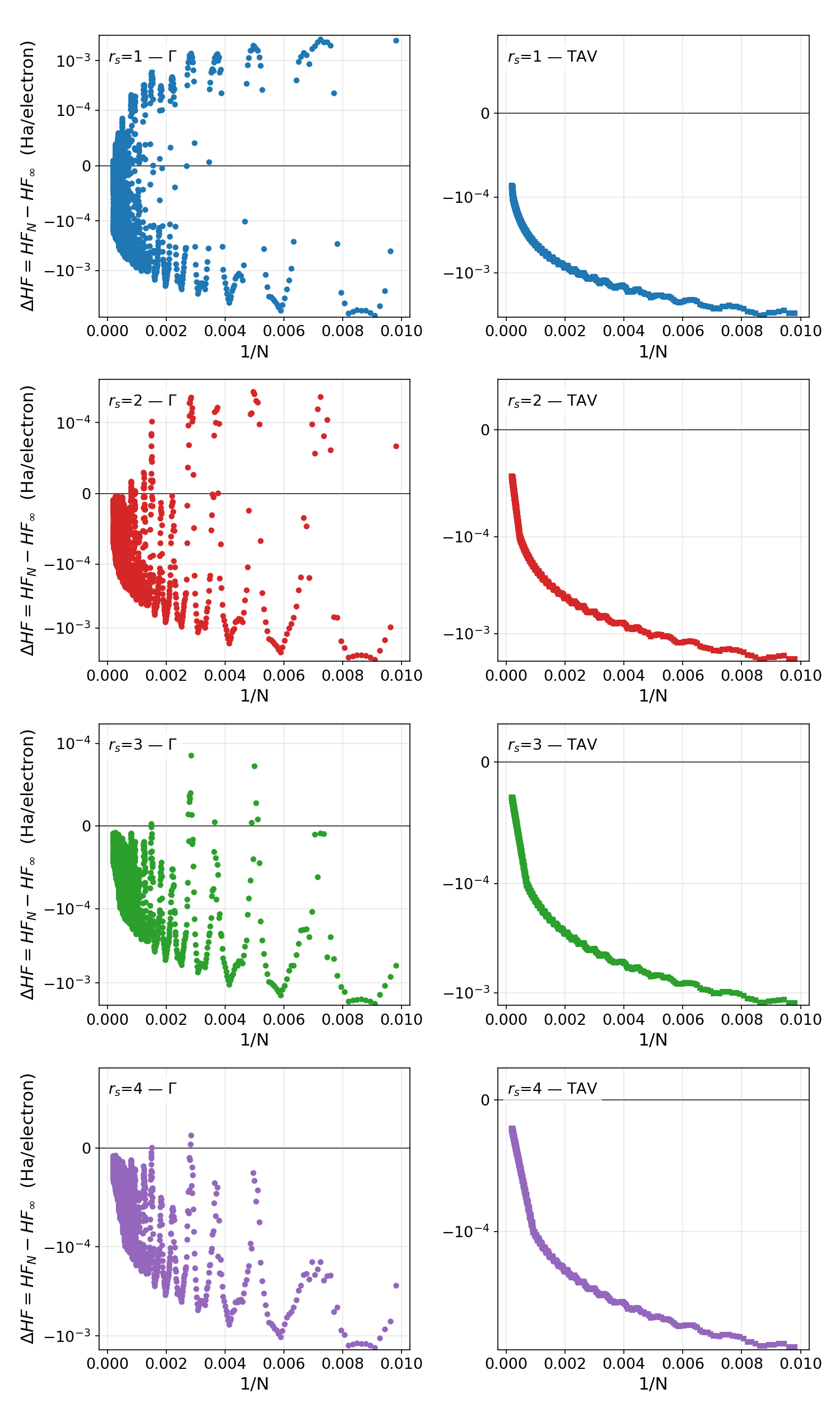}
    \caption{Finite-cell HF energy error, $\Delta E^{\rm HF}=E_N^{\rm HF}-E_\infty^{\rm HF}$ (Ha/electron), as a function of $1/N$ for the unpolarized ($\zeta=0$) 2D UEG at $r_s=1,2,3,4$, using hexagonal simulation cells, $N=102$--$5000$. $\Gamma$-point (left column) and TA (right column) results are shown separately, one row per density. The $\Gamma$-point results oscillate in sign; the TA results remain single-signed throughout.}
    \label{fig:HF}
\end{figure}

\begin{figure}[t]
    \centering
    \includegraphics[width=1.0\linewidth]{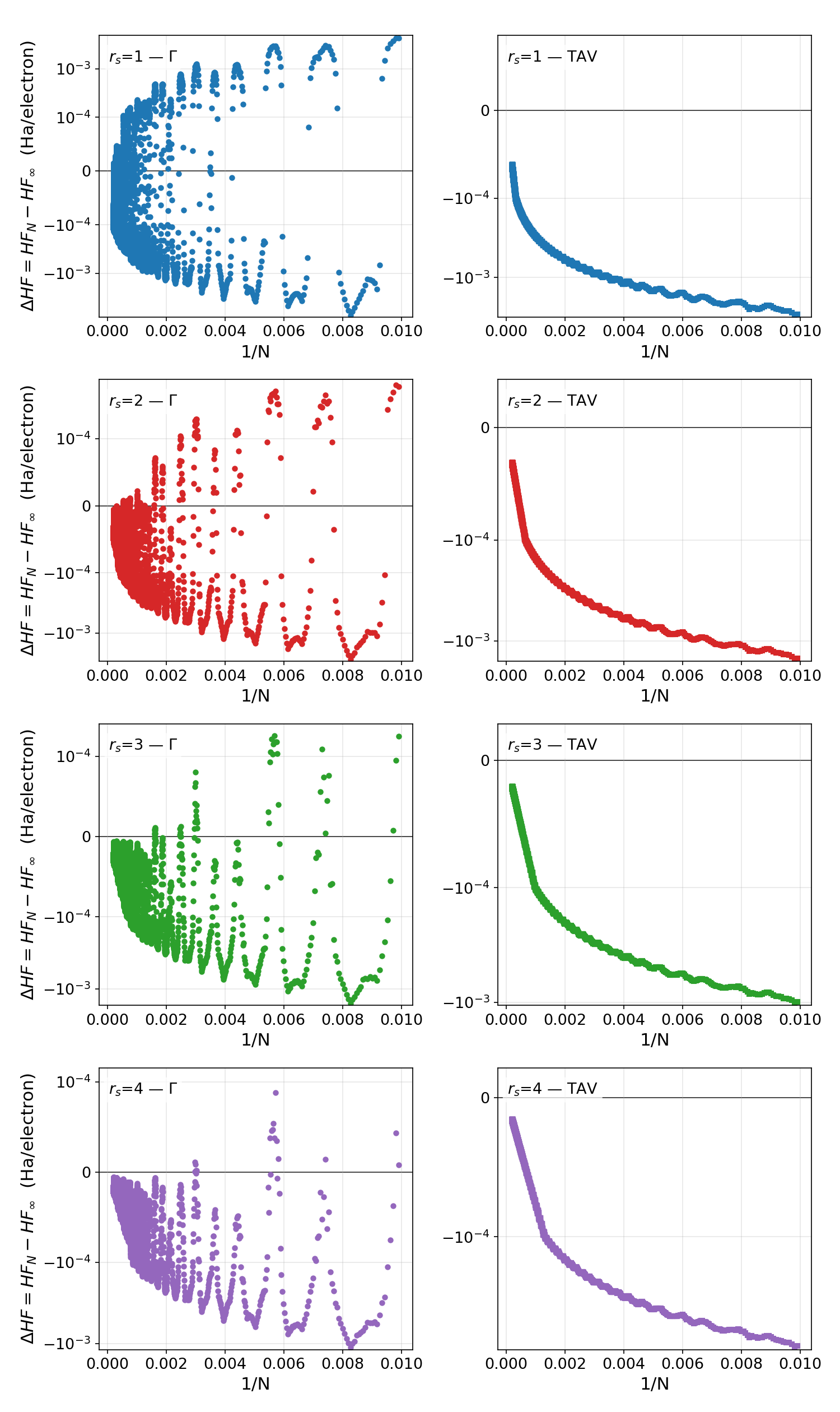}
    \caption{As Fig.~\ref{fig:HF}, for the fully spin-polarized ($\zeta=1$) 2D UEG.}
    \label{fig:HF_zeta1}
\end{figure}

We define the finite-cell HF error as
\begin{equation}
\Delta E^{\rm HF}(N) = E_N^{\rm HF} - E_\infty^{\rm HF},
\end{equation}
and identify Good-$\Gamma$ (GG) particle numbers as those $N$, within a chosen range, for which $|\Delta E^{\rm HF}(N)|$ is minimized. All results in this section use $N>100$ throughout. Figures~\ref{fig:HF} and \ref{fig:HF_zeta1} show $\Delta E^{\rm HF}(N)$ for $\Gamma$-point and TA HF calculations at $r_s=1,2,3,4$, over $N=102$ to $5000$, for $\zeta=0$ and $\zeta=1$ respectively obtained using $10^4$ twists. In both cases, the $\Gamma$-point results show the large shell-filling oscillations that are well known for $\Gamma$-point calculations of the UEG \cite{Lin01,Foulkes01}, while the TA results vary smoothly with $N$ and do not minimize $|\Delta E^{\rm HF}(N)|$ at any particular $N$. The oscillation itself is not new; what we examine here is its decomposition into kinetic- and exchange-energy components, and the consequences of that decomposition for the particular $N$ at which the total error nearly vanishes.

\subsection{Cancellation between $\Delta$KE and $\Delta$EX at Good-$\Gamma$ numbers}
\label{sec:cancellation}

At Good-$\Gamma$ numbers, $\Delta KE(N)$ and $\Delta EX(N)$ are individually large and opposite in sign, canceling in the total $\Delta E^{\rm HF}(N)$.  We tested this directly at every GG number identified across $N=102$--$5000$, at all four densities and both polarizations and found that $\Delta KE(N)$ and $\Delta EX(N)$ are opposite in sign and individually $5\times$ to more than $10^4\times$ larger in magnitude than their sum, with the smaller ratios typical and the largest reflecting a small number of near-exact cancellations.

\begin{table}[h]
\centering
\begin{tabular}{c c c}
\hline\hline
$\zeta$ & $r_s$ & median $\Delta EX/\Delta KE$ \\
\hline
0 & 1 & $-0.727$ \\
0 & 2 & $-1.454$ \\
0 & 3 & $-2.191$ \\
0 & 4 & $-2.909$ \\
\hline
1 & 1 & $-0.654$ \\
1 & 2 & $-1.309$ \\
1 & 3 & $-1.965$ \\
1 & 4 & $-2.617$ \\
\hline\hline
\end{tabular}
\caption{Median $\Delta EX(N)/\Delta KE(N)$ over $N=102$--$5000$, restricted at each density to the half of that range with the larger $|\Delta KE(N)|$ (a rank-based, not absolute, cut, chosen so the same fraction and count of points enters the median at every density -- an absolute cut biases the sample toward progressively fewer, smaller-$N$ points as $r_s$ increases, since $\Delta KE$ shrinks with $r_s$). The ratio is negative at every density and both polarizations tested, and its magnitude increases with $r_s$ in close to direct proportion (1:2:3:4) at both polarizations.}
\label{tab:ratio}
\end{table}

Table~\ref{tab:ratio} shows that $\Delta EX(N)/\Delta KE(N)$ is negative at every density and both polarizations, confirming that the two errors are consistently opposite in sign, not just occasionally so. Its magnitude grows with $r_s$ in close to exact proportion (1:2:3:4 at both polarizations), and is somewhat larger at $\zeta=0$ than at $\zeta=1$ for a given $r_s$. We report the median rather than the mean because $\Delta KE(N)$ oscillates through zero, so a small number of points close to a zero crossing produce near-singular ratios that can dominate a mean (at $r_s=1$, $\zeta=0$, for example, the mean of the same subset has the wrong sign) but leave the median unaffected.

\subsection{One-body structure beyond the cancellation ratio}
\label{sec:hf_structure}

Three further HF-level properties, established from the $N=102$--$5000$ dataset, sharpen this picture. First, it is the kinetic-energy error, not the exchange-energy error, that qualitatively distinguishes $\Gamma$-point from TA behavior: over this range, TA's $\Delta KE(N)$ and $\Delta EX(N)$ never change sign, and neither does the $\Gamma$-point $\Delta EX(N)$; only $\Gamma$-point $\Delta KE(N)$ changes sign, repeatedly, 58 times for $\zeta=0$ and 81 for $\zeta=1$, identically at all four densities in each case. Second, at $N\geq4000$ the $\Gamma$-point kinetic-energy error is $27$--$28\times$ larger than TA's at both polarizations, while its exchange-energy error is only $0.91$--$0.96\times$ TA's; TA's advantage over a generic $\Gamma$-point calculation is thus overwhelmingly a kinetic-energy effect, and for exchange alone a single well-chosen $\Gamma$-point calculation remains essentially as converged as TA even at the largest $N$ studied. These figures hold to the precision quoted regardless of density or polarization, indicating they are intrinsic to the $\Gamma$-point occupation rather than incidental to any one case.

\begin{figure}[t]
    \centering
    \includegraphics[width=1.0\linewidth]{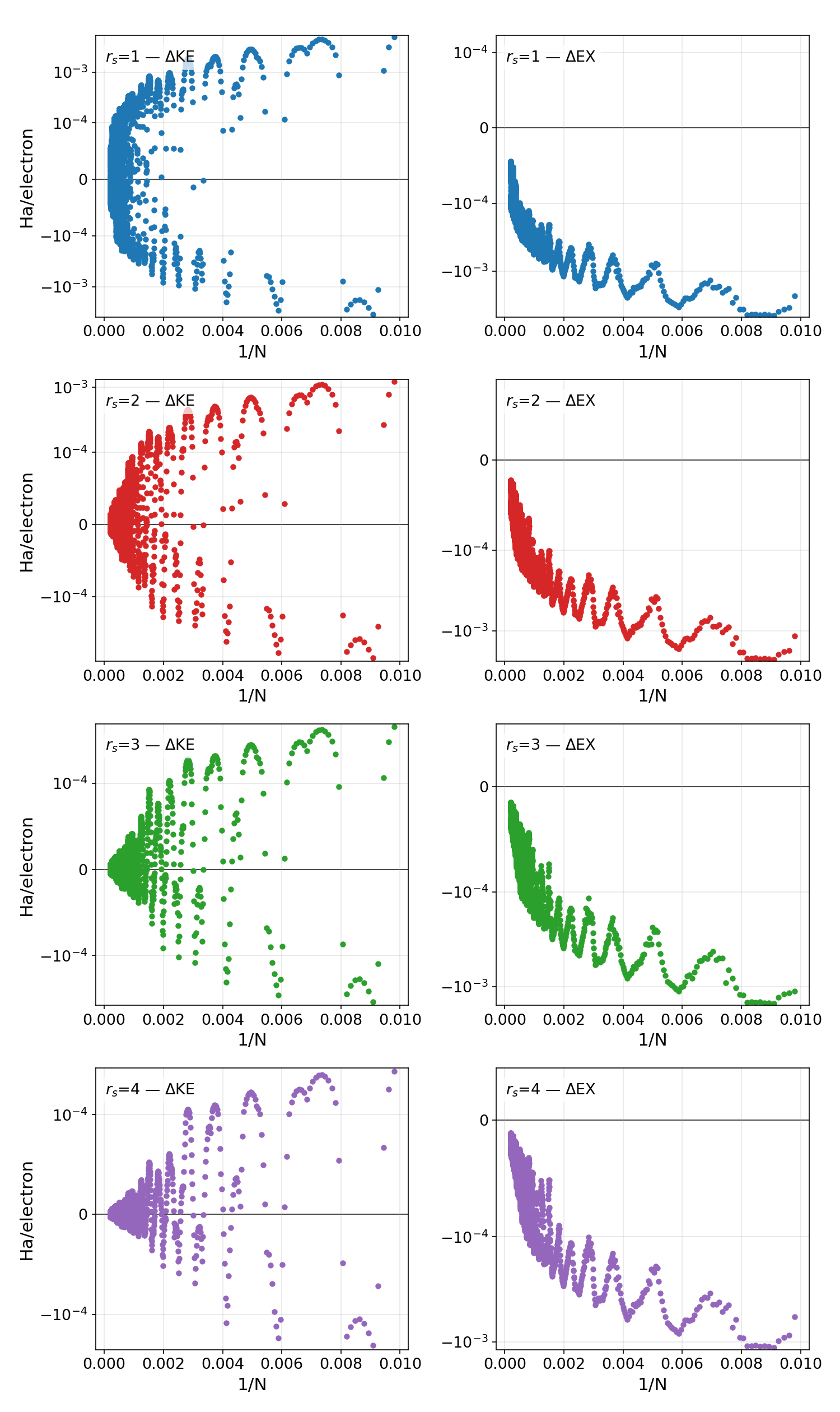}
    \caption{$\Gamma$-point $\Delta KE(N)$ (left) and $\Delta EX(N)$ (right) vs.\ $1/N$, unpolarized ($\zeta=0$) 2D UEG, $r_s=1,2,3,4$ (one row per density), $N=102$--$5000$. $\Delta EX(N)$ decays smoothly and is single-signed at every density; $\Delta KE(N)$ oscillates through zero repeatedly, with an envelope that decays much more slowly.}
    \label{fig:components}
\end{figure}

\begin{figure}[t]
    \centering
    \includegraphics[width=1.0\linewidth]{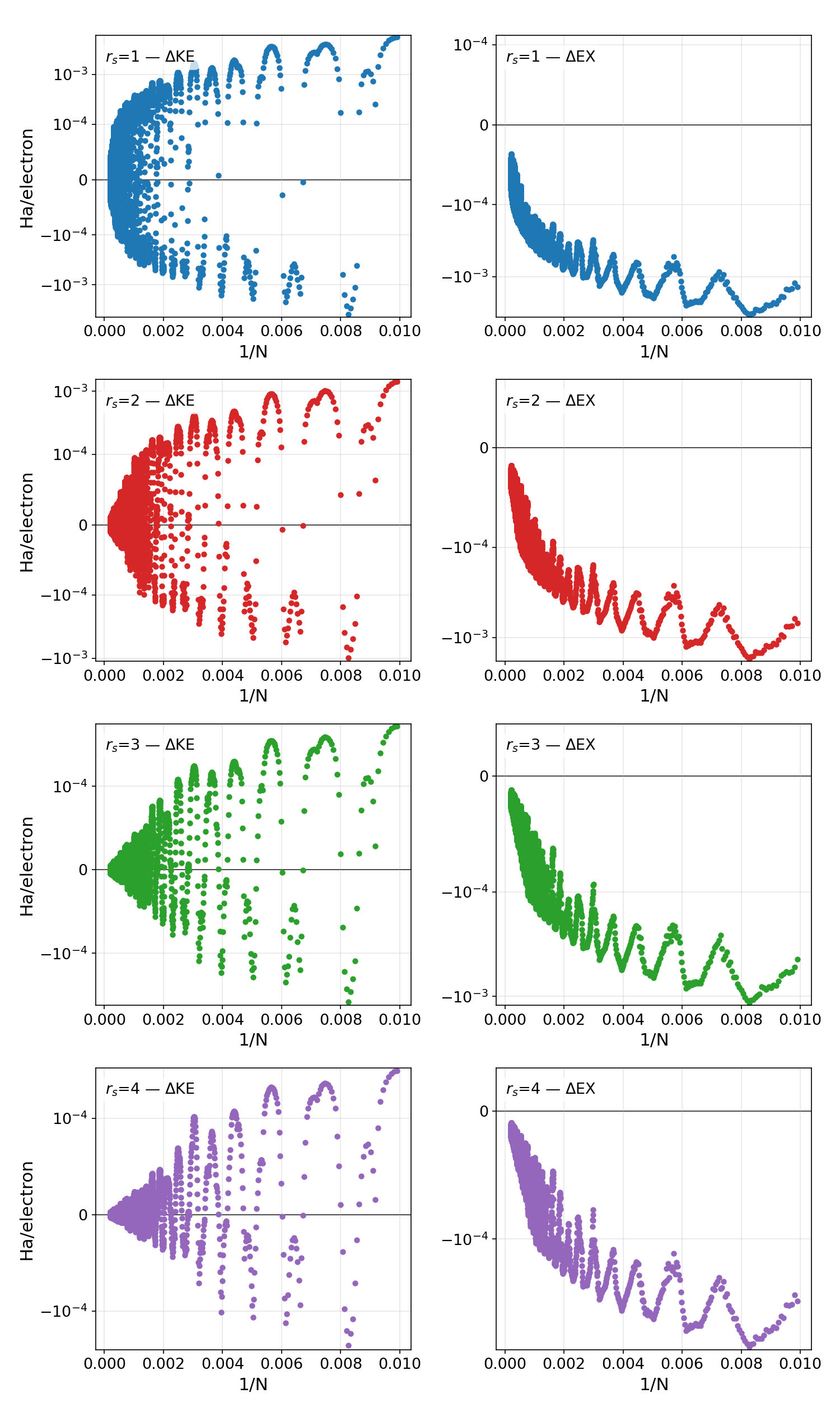}
    \caption{As Fig.~\ref{fig:components}, for the fully spin-polarized ($\zeta=1$) 2D UEG.}
    \label{fig:components_zeta1}
\end{figure}

\begin{figure}[t]
    \centering
    \includegraphics[width=1.0\linewidth]{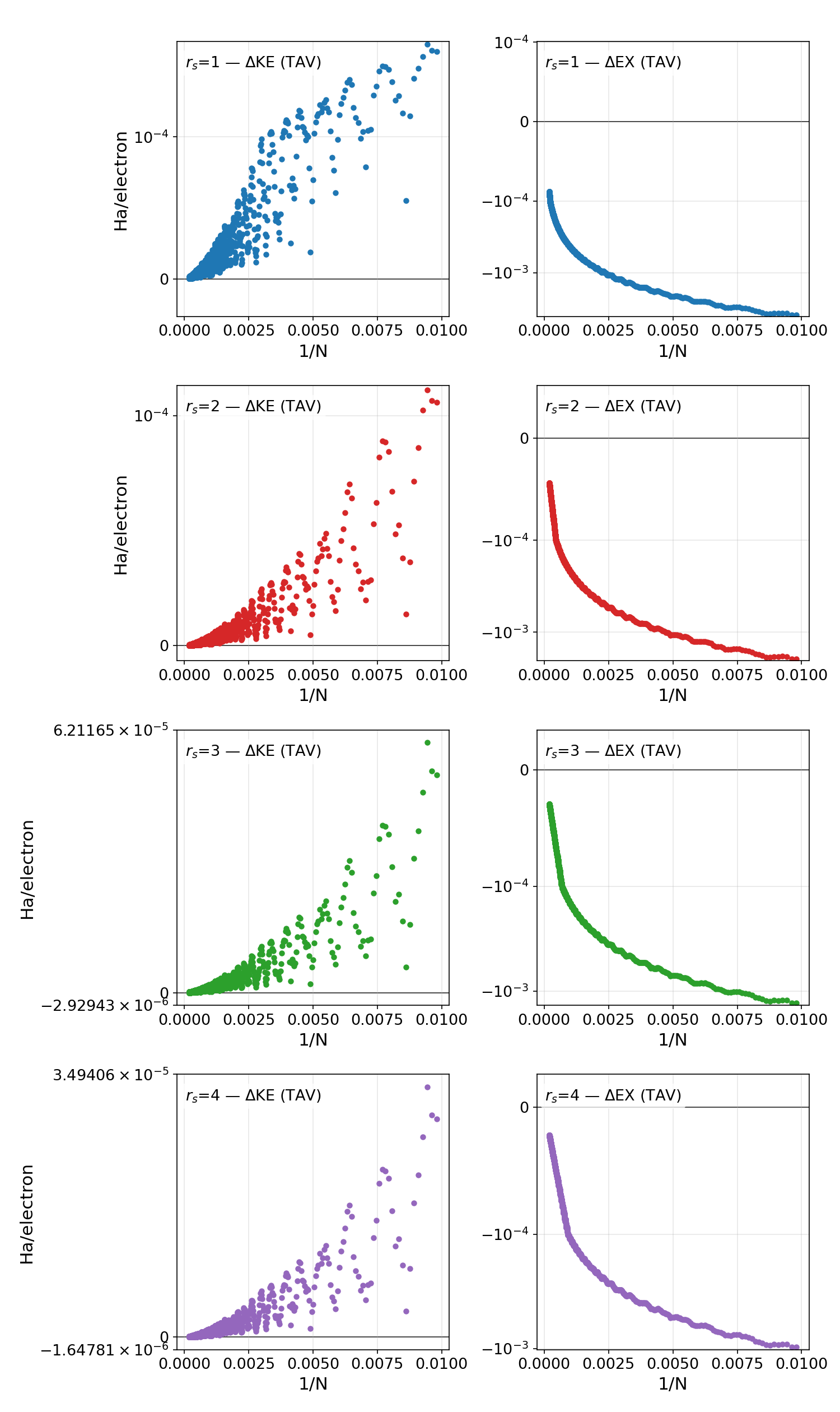}
    \caption{As Fig.~\ref{fig:components}, but for the twist-averaged $\Delta KE_{\rm TAV}(N)$ and $\Delta EX_{\rm TAV}(N)$ rather than the $\Gamma$-point values, unpolarized ($\zeta=0$) 2D UEG.}
    \label{fig:components_tav}
\end{figure}

Figures~\ref{fig:components} and \ref{fig:components_zeta1} show $\Delta KE(N)$ and $\Delta EX(N)$ directly, for $\zeta=0$ and $\zeta=1$ respectively. The contrast is visually immediate in both cases, with $\Delta EX(N)$ tracing a single, smoothly decaying, single-signed curve at every density while $\Delta KE(N)$ oscillates through zero with a slowly decaying envelope. Figure~\ref{fig:components_tav} shows the same two quantities for the TA energies, both of which decay smoothly and remain single-signed throughout, with none of the $\Gamma$-point oscillation visible above, which is a direct visual confirmation of the sign-change counts just described.

Third, the set of GG-$N$ is density-dependent, meaning no single $N$ ranks among the best Good-$\Gamma$ candidates at all four densities simultaneously, at either polarization, and the overlap between the GG sets at two densities is larger the closer those densities are (the overlap counts are tabulated in the Supplemental Material \cite{SM}). A practical consequence is that GG numbers must be identified separately at each $r_s$ and $\zeta$ from the $\Gamma$-point HF scan, which is inexpensive; there is no universal GG set.

\subsection{Per-twist structure: testing the sign asymmetry directly}
\label{sec:per_twist}

\begin{table}[h]
\centering
\begin{tabular}{c c c c c}
\hline\hline
\footnotesize $N$ & type & $\Delta E^{\rm HF}_\Gamma$ & $\Delta KE$ \% pos.\ (5k) & $\Delta KE$ \% pos.\ (10k) \\
\hline
104 & GG & $-4.00\times10^{-4}$ & 81.0\% & 80.5\% \\
128 & GG & $-2.87\times10^{-4}$ & 82.6\% & 82.3\% \\
158 & GG & $-1.23\times10^{-5}$ & 90.4\% & 91.0\% \\
190 & GG & $+2.61\times10^{-4}$ & 51.8\% & 51.8\% \\
250 & arbitrary & $-1.79\times10^{-3}$ & 72.0\% & 72.3\% \\
300 & arbitrary & $-1.84\times10^{-3}$ & 75.0\% & 75.6\% \\
400 & arbitrary & $-2.11\times10^{-3}$ & 53.9\% & 54.8\% \\
500 & arbitrary & $-1.50\times10^{-3}$ & 63.1\% & 63.7\% \\
\hline\hline
\end{tabular}
\caption{Per-twist statistics of $\Delta KE$ and $\Delta EX$ at $r_s=1$, $\zeta=0$, for four Good-$\Gamma$ and four arbitrary $N$, at 5000 and 10000 twists per $N$. $\Delta EX$ is negative at every individual twist at every $N$; $\Delta KE$ is positive for a majority of twists at every $N$. The fraction of positive $\Delta KE$ changes by less than one percentage point between 5000 and 10000 twists, consistent with statistical noise, so it is converged at 5000 twists.}
\label{tab:per_twist}
\end{table}

\begin{figure}[t]
\centering
\includegraphics[width=0.85\linewidth]{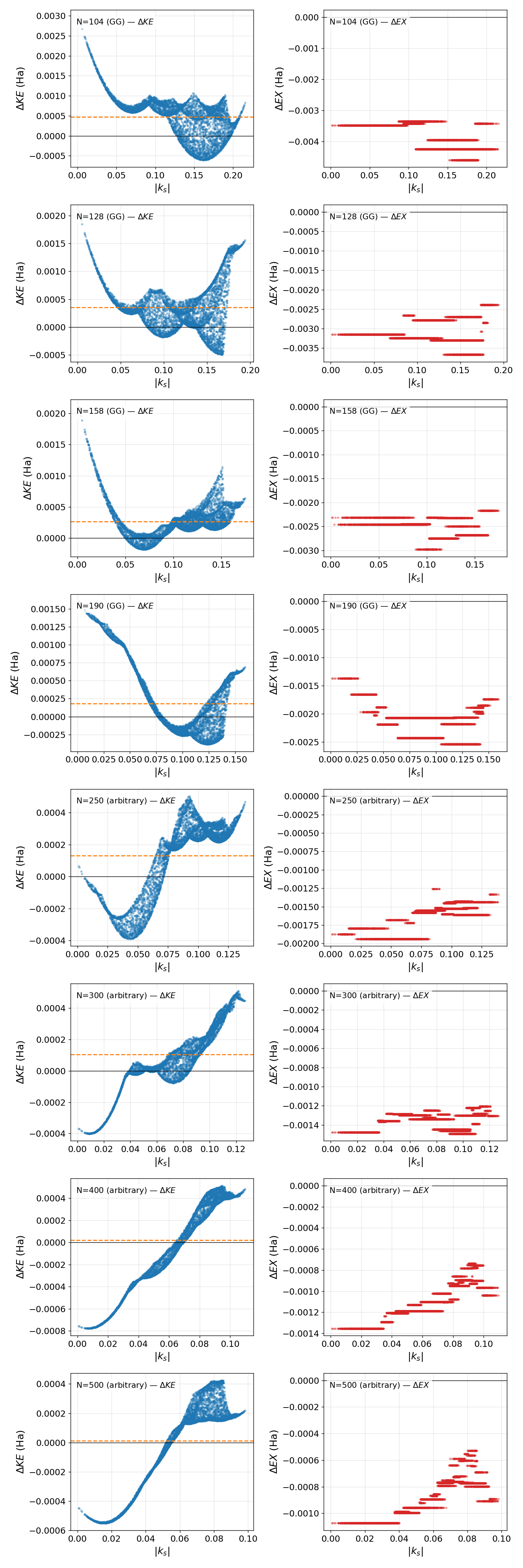}
\caption{$\Delta KE$ (left) and $\Delta EX$ (right) vs.\ twist-vector magnitude $|k_s|$, $r_s=1$, $\zeta=0$, 5000 twists per panel, one row per $N$ (top four: Good-$\Gamma$ $N$ with QMC data; bottom four: $N$ chosen without reference to $\Delta E^{\rm HF}$, within the range most relevant to QMC calculations). The dashed line in each $\Delta KE$ panel marks the twist-averaged value.}
\label{fig:twist_vs_k}
\end{figure}

Two results (Table~\ref{tab:per_twist}) are robust across the eight $N$ considered, including the four chosen with no reference to $|\Delta E^{\rm HF}(N)|$, and are therefore properties of the TA distribution itself rather than artifacts of the Good-$\Gamma$ selection: (i) $\Delta EX$ is negative at every individual twist, not merely on average -- a stronger statement than the twist-averaged result of Sec.~\ref{sec:hf_structure}; and (ii) $\Delta KE$ is positive for a majority of twists at every $N$ tested, confirming that the positive sign of the twist-averaged $\Delta KE$ (dashed line, Fig.~\ref{fig:twist_vs_k}) reflects a genuine majority rather than a small number of large outliers. Figure~\ref{fig:twist_vs_k} further shows that $\Delta KE$ and $\Delta EX$ depend on the twist vector itself very differently at the two kinds of $N$ studied.

\subsection{Thermodynamic-limit extrapolation within the QMC-relevant range}
\label{sec:range_100_500}

We restrict this comparison to $100<N<500$, the range within which QMC calculations in this work are actually performed. For each density, we linearly extrapolate ($1/N$) the TA-HF energy using all available $N$ in this range, and separately extrapolate the $\Gamma$-point HF energy using four Good-$\Gamma$ $N$ selected to have $|\Delta E^{\rm HF}_\Gamma|$ as small as possible subject to a minimum spacing of 50 in $N$ between any two selected points, chosen so the linear fit is not dominated by nearly-degenerate points. The four selected $N$ are $\{214,290,372,468\}$, $\{208,266,368,454\}$, $\{138,204,274,350\}$, and $\{202,274,354,450\}$ for $r_s=1,2,3,4$ respectively; the minimum gap between consecutive selected $N$ is 58--96, comfortably spread across the 400-wide window at every density. Both extrapolations are compared against the exact analytic thermodynamic-limit value.

\begin{table}[h]
\centering
\begin{tabular}{c c c}
\hline\hline
$r_s$ & TAV (all $N$) & GG-$N$ (4 points) \\
\hline
1 & $-0.113$ & $+0.146$ \\
2 & $-0.042$ & $+0.005$ \\
3 & $-0.025$ & $+0.025$ \\
4 & $-0.017$ & $-0.047$ \\
\hline\hline
\end{tabular}
\caption{Difference (mHa) between the linearly extrapolated ($1/N$) HF energy and the exact analytic value, $100<N<500$, $\zeta=0$.}
\label{tab:range_100_500}
\end{table}

\begin{figure*}[t]
\centering
\begin{tabular}{cc}
     \includegraphics[width=0.5\linewidth]{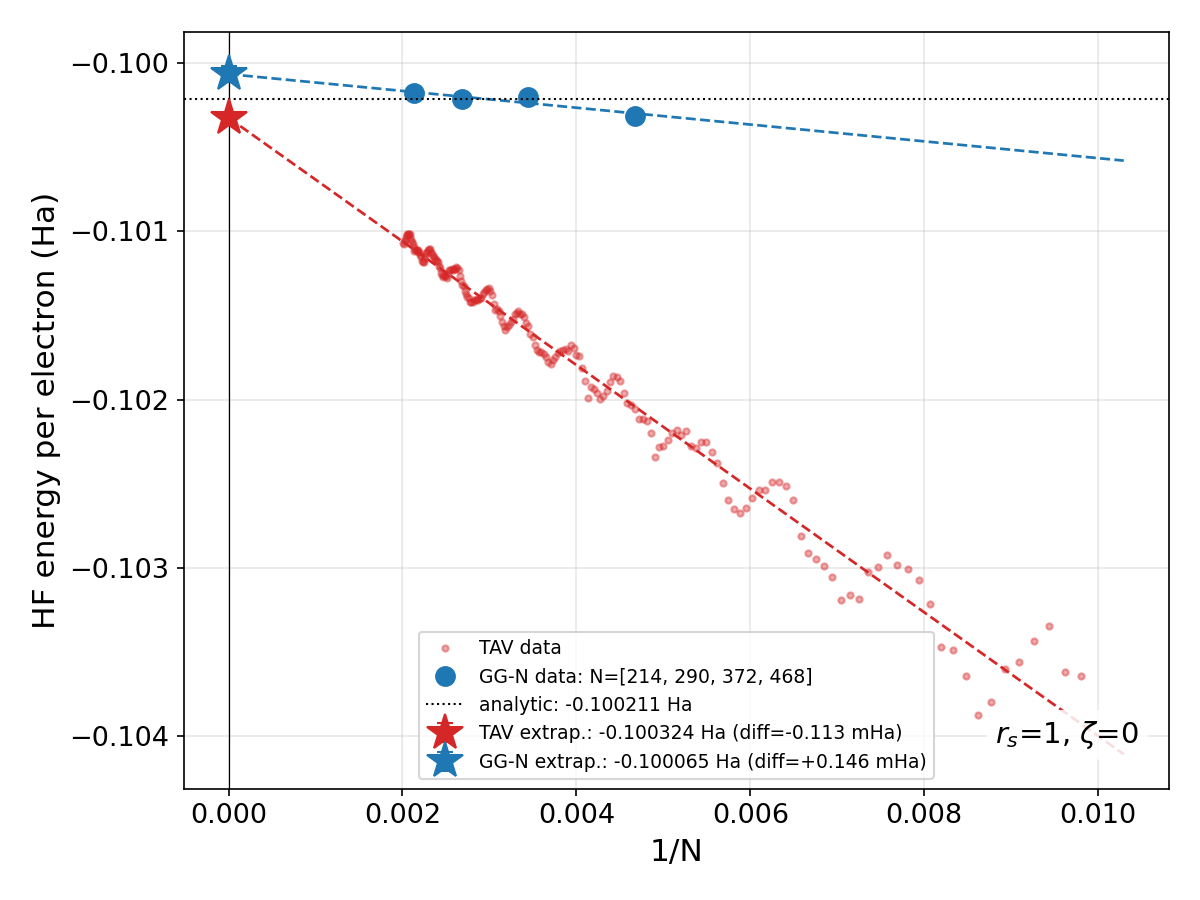} & 
     \includegraphics[width=0.5\linewidth]{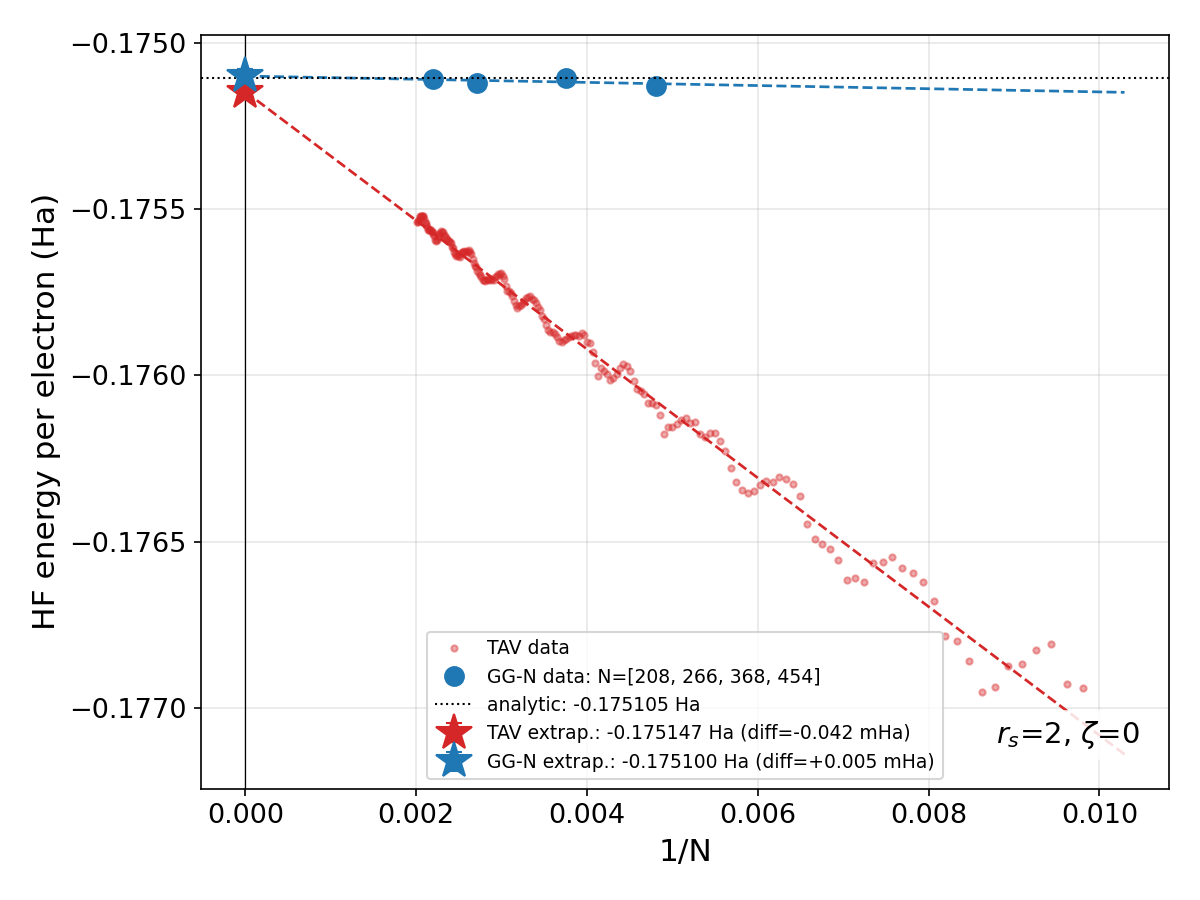}\\
     \includegraphics[width=0.5\linewidth]{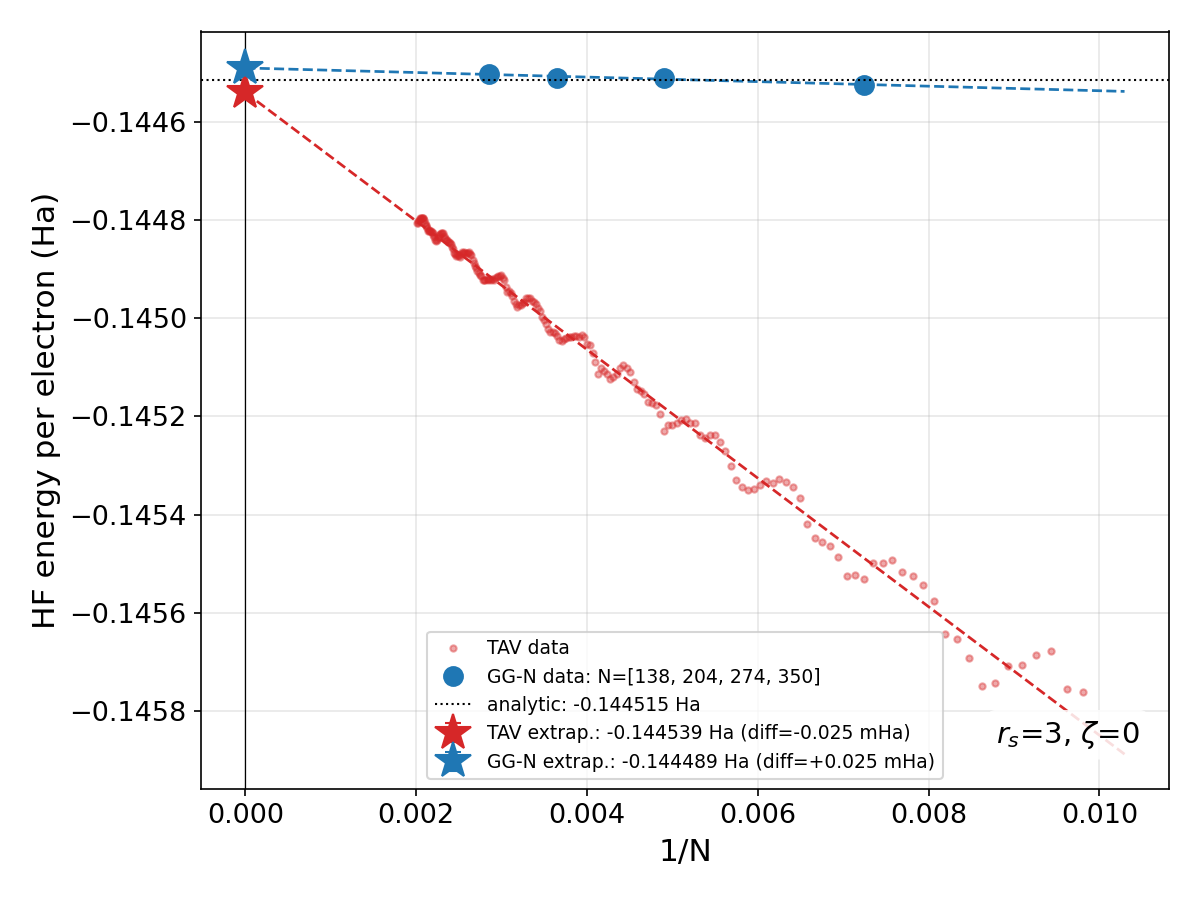} &
     \includegraphics[width=0.5\linewidth]{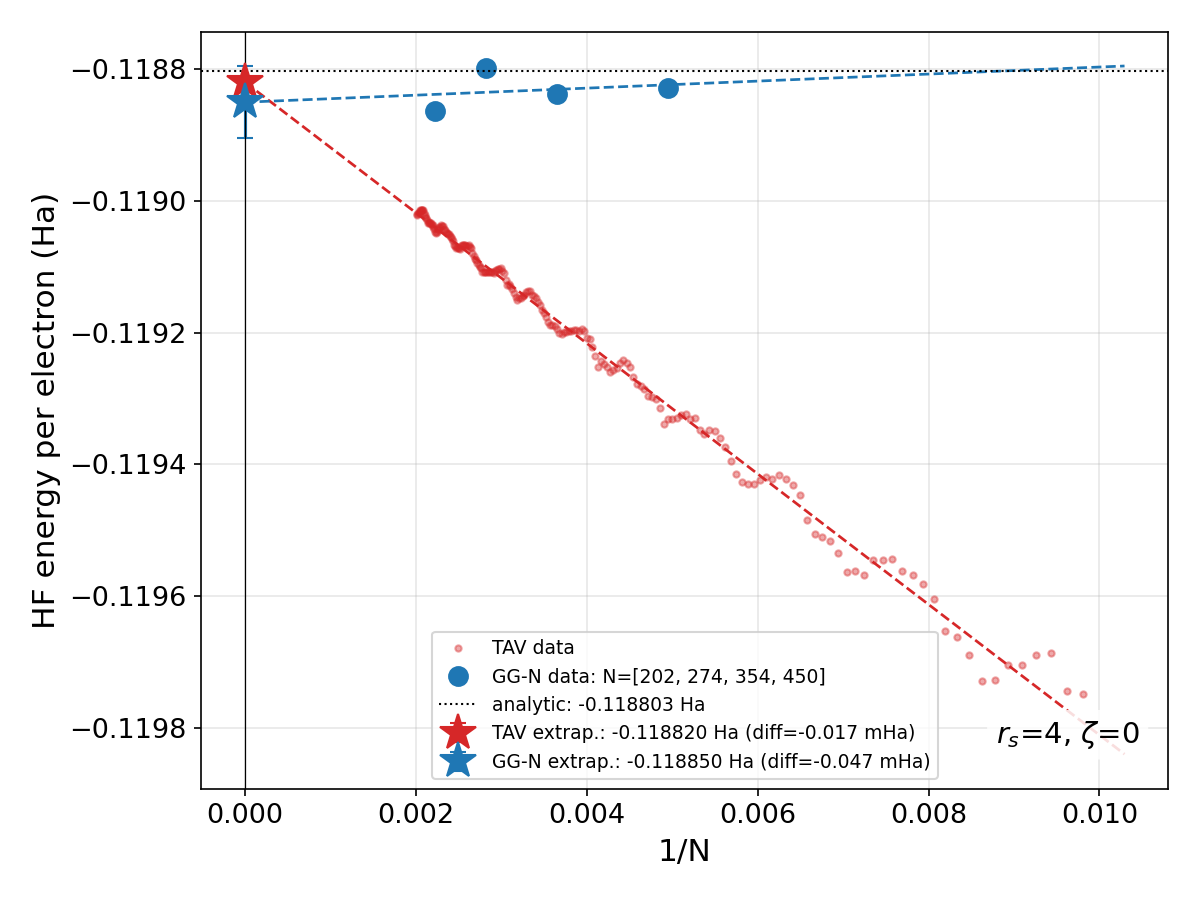} \\
\end{tabular}
\caption{HF energy per electron vs.\ $1/N$, $r_s=1,2,3,4$, $\zeta=0$, $100<N<500$: twist-averaged energy (all available $N$) and $\Gamma$-point energy at four well-spaced Good-$\Gamma$ $N$, each linearly extrapolated to $1/N=0$ and compared against the exact analytic value (dotted line).}
\label{fig:range100500_rs1}
\end{figure*}

Both extrapolations agree with the analytic value to well within 0.2\,mHa at every density (Table~\ref{tab:range_100_500}, Fig.~\ref{fig:range100500_rs1}), and neither is uniformly better than the other: GG-$N$ is more accurate at $r_s=2$, comparably accurate at $r_s=3$, and less accurate than TAV at $r_s=1$ and $r_s=4$. This shows that, once the Good-$\Gamma$ $N$ are chosen to be well separated in $1/N$ rather than selected by rank alone (Sec.~\ref{sec:hf_structure}), a handful of $\Gamma$-point calculations can extrapolate to the thermodynamic limit about as reliably as a full TA sequence over the same range, at the HF level.

\section{Quantum Monte Carlo}
\label{sec:qmc}

We test whether the HF-level GG criterion, shown in Sec.~\ref{sec:hf}, provides a practical route to the DMC thermodynamic limit (TL). Throughout this section we compare two DMC sequences evaluated at the same Good-$\Gamma$ particle numbers, including the $\Gamma$-point DMC energies themselves (``GG''), and the twist-averaged DMC energies at those same $N$ (``TAV-GG'', twist-averaged at Good-$\Gamma$ $N$). The latter serves as the reference against which the $\Gamma$-point sequence is judged, since it removes the one-body shell error by averaging rather than by cancellation. The result is that a GG-based DMC sequence extrapolates to the TL in close agreement with the TAV-GG extrapolation, at every density and polarization tested (Sec.~\ref{sec:extrapolation}), while requiring substantially fewer independent DMC calculations to obtain (Sec.~\ref{sec:discussion}). This does not mean single-$N$ GG accuracy is guaranteed at the DMC level, as the correlated-energy finite-size error is expected to be dominated by many-body effects that a purely one-body criterion does not address, and we do not find that it is guaranteed. What matters for a converged benchmark, however, is the extrapolated result, and that is where GG performs reliably. A direct test isolating the one-body contribution to the correlated-energy error, contrasting GG against generic (``Bad-$\Gamma$'') particle numbers known to be poor at the one-body level, is given in the Supplemental Material \cite{SM}.

\subsection{Resolving the opposite-slope puzzle}
\label{sec:opposite_slope}

The GG and TA-DMC sequences approach the TL from opposite sides as functions of $1/N$. Our results show that TA-DMC energies satisfy $E_{\rm TAV}(N)-E(\infty)<0$ (the finite cell overbinds), while GG DMC energies satisfy $E_{\rm GG}(N)-E(\infty)>0$. This sign reflects an over correction, as the many-body finite-size correction beyond the one-body error is generically positive \cite{Chiesa06,Kent99}, while the one-body cancellation at GG numbers leaves no reason for the corresponding interacting-system one-body bias to vanish at the same $N$. When this residual one-body bias is larger in magnitude than, and opposite in sign to, the true positive many-body correction, it flips the net sign of the observed error, producing the anomalous $E_{\rm GG}(N)-E(\infty)>0$ trend.

\begin{figure}[t]
\centering
\includegraphics[width=1.0\linewidth]{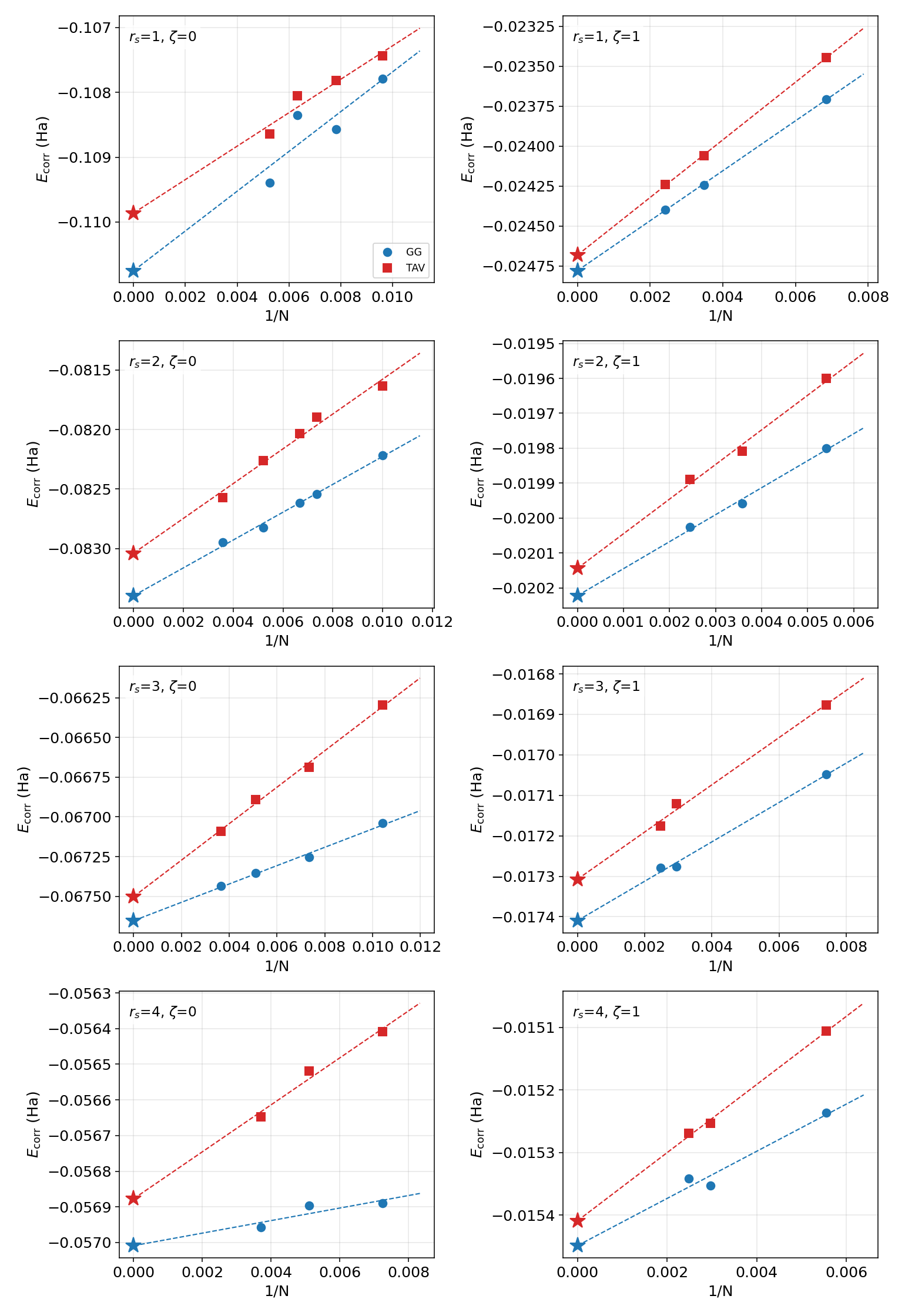}
\caption{Correlation energy $E_{\rm corr}=E_{\rm DMC}-E_{\rm HF}$ vs.\ $1/N$, computed separately along the GG and TAV sequences, all four densities and both polarizations. In every one of the 16 panel/sequence combinations, a linear fit to $E_{\rm corr}(N)$ has a positive slope.}
\label{fig:ecorr_all}
\end{figure}

The positive sign of the many-body correction, established in the literature \cite{Chiesa06,Kent99}, can also be tested directly from our own data, without relying on an external correction scheme. We define the correlation energy as $E_{\rm corr}(N) = E_{\rm DMC}(N) - E_{\rm HF}(N)$, computed separately along the GG and TAV sequences using each sequence's own $\Gamma$-point or TA-HF energy at the same $N$, and examine its dependence on $1/N$ (Fig.~\ref{fig:ecorr_all}, Table~\ref{tab:ecorr}). At every density, both polarizations, and both sequences -- 16 independent tests in total -- a linear fit to $E_{\rm corr}(N)$ vs.\ $1/N$ has a positive slope, i.e.\ the trend is toward a more negative (more strongly bound) correlation energy as $N$ increases, with no exceptions. Two of the sixteen raw sequences (each with only 3--4 available $N$) show a small point-to-point deviation from strict monotonicity, plausibly ordinary scatter given the small sample size; nonetheless, every individual finite-$N$ data point in all sixteen sequences is less negative than its own linearly extrapolated $N\to\infty$ value, directly confirming a positive correlation-energy finite-size error using only our own raw, uncorrected DMC and HF data.

\begin{table}[h]
\centering
\begin{tabular}{c c c c}
\hline\hline
$\zeta$ & $r_s$ & slope, GG (mHa) & slope, TAV (mHa) \\
\hline
0 & 1 & 306.7 & 257.9 \\
0 & 2 & 116.8 & 146.1 \\
0 & 3 & 57.8  & 114.8 \\
0 & 4 & 17.6  & 65.9  \\
\hline
1 & 1 & 156.3 & 180.3 \\
1 & 2 & 77.3  & 99.1  \\
1 & 3 & 48.7  & 58.3  \\
1 & 4 & 37.7  & 54.4  \\
\hline\hline
\end{tabular}
\caption{Slope $b$ of the linear fit $E_{\rm corr}(N) = E_{\rm corr}(\infty) + b/N$ to the GG and TAV correlation-energy sequences of Fig.~\ref{fig:ecorr_all}. The slope is positive at every density, polarization, and sequence tested (16/16), confirming a positive correlation-energy finite-size error directly from our own data.}
\label{tab:ecorr}
\end{table}

\subsection{Linearity and thermodynamic-limit extrapolation}
\label{sec:extrapolation}

\begin{figure*}
    \centering
    \begin{tabular}{cc}
             \includegraphics[width=0.95\linewidth]{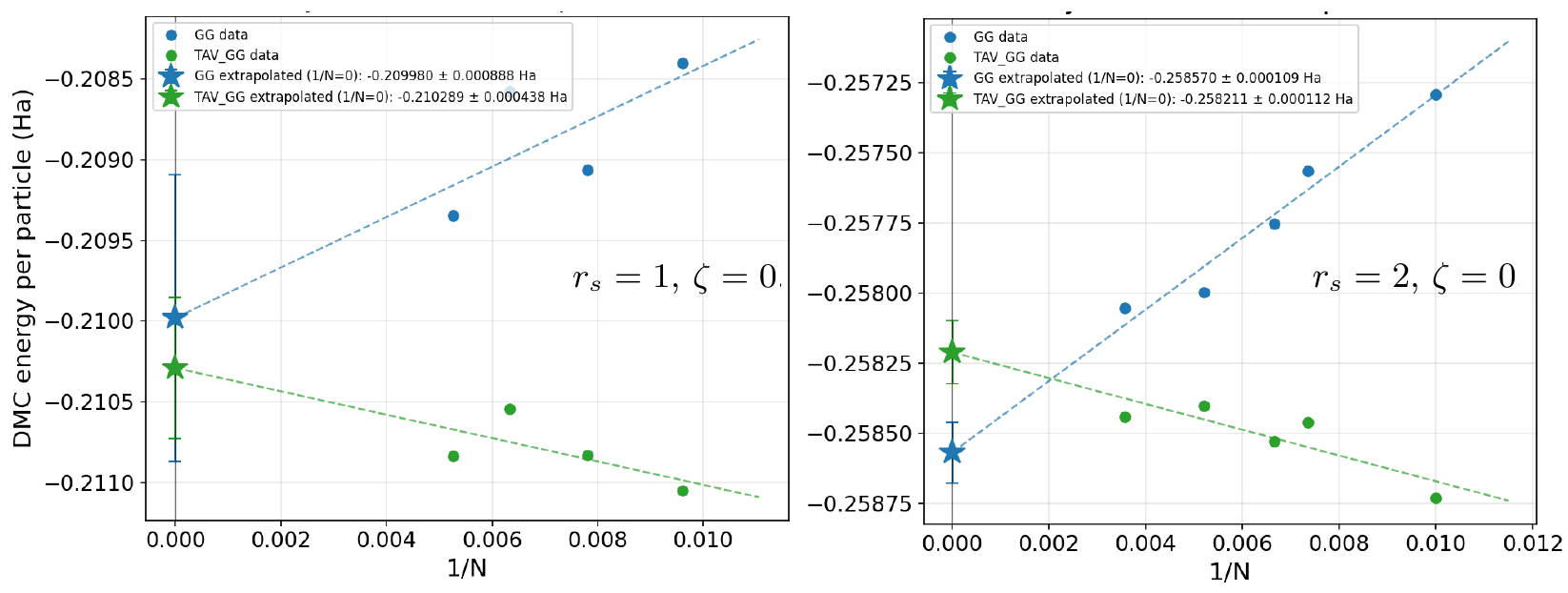}  \\
             \includegraphics[width=0.95\linewidth]{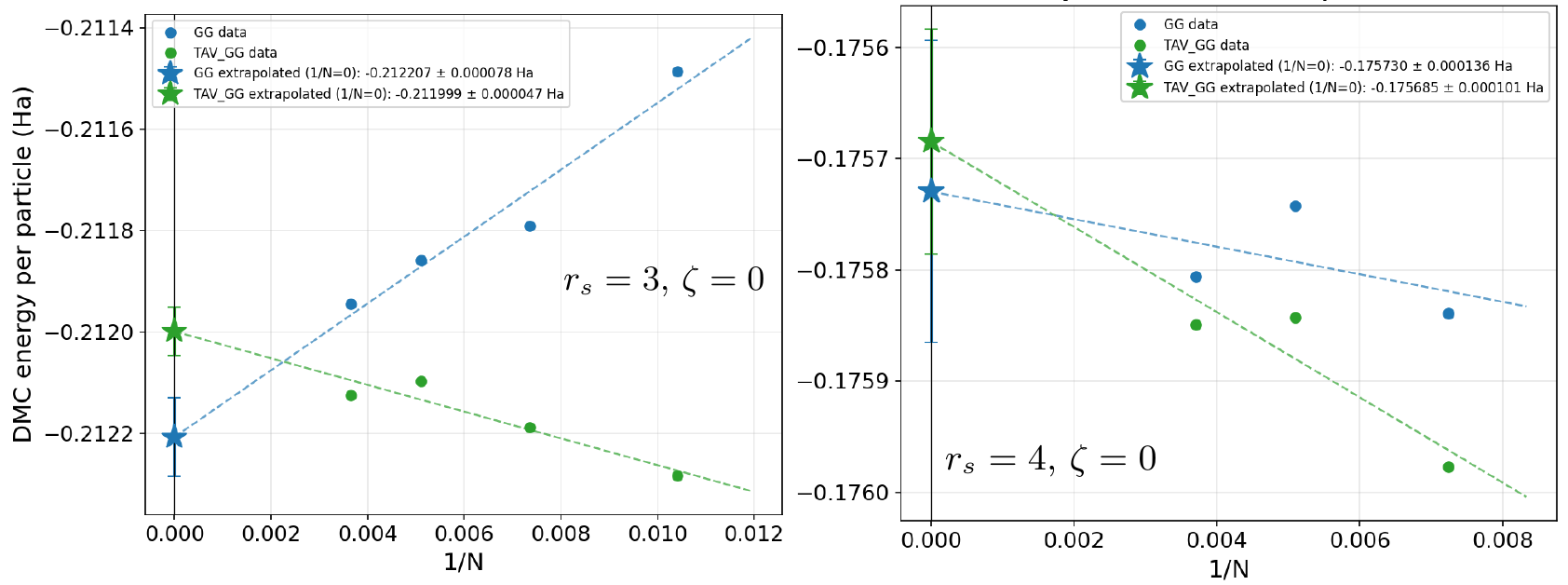}  \\ 
    \end{tabular}
    \caption{Linear extrapolation of GG and TAV-GG DMC energies to $1/N=0$, 2D UEG at $r_s=1,2,3,4$, $\zeta=0$.}
    \label{fig:rs1_extrap}
\end{figure*}

\begin{figure*}
    \centering
    \begin{tabular}{cc}
             \includegraphics[width=0.95\linewidth]{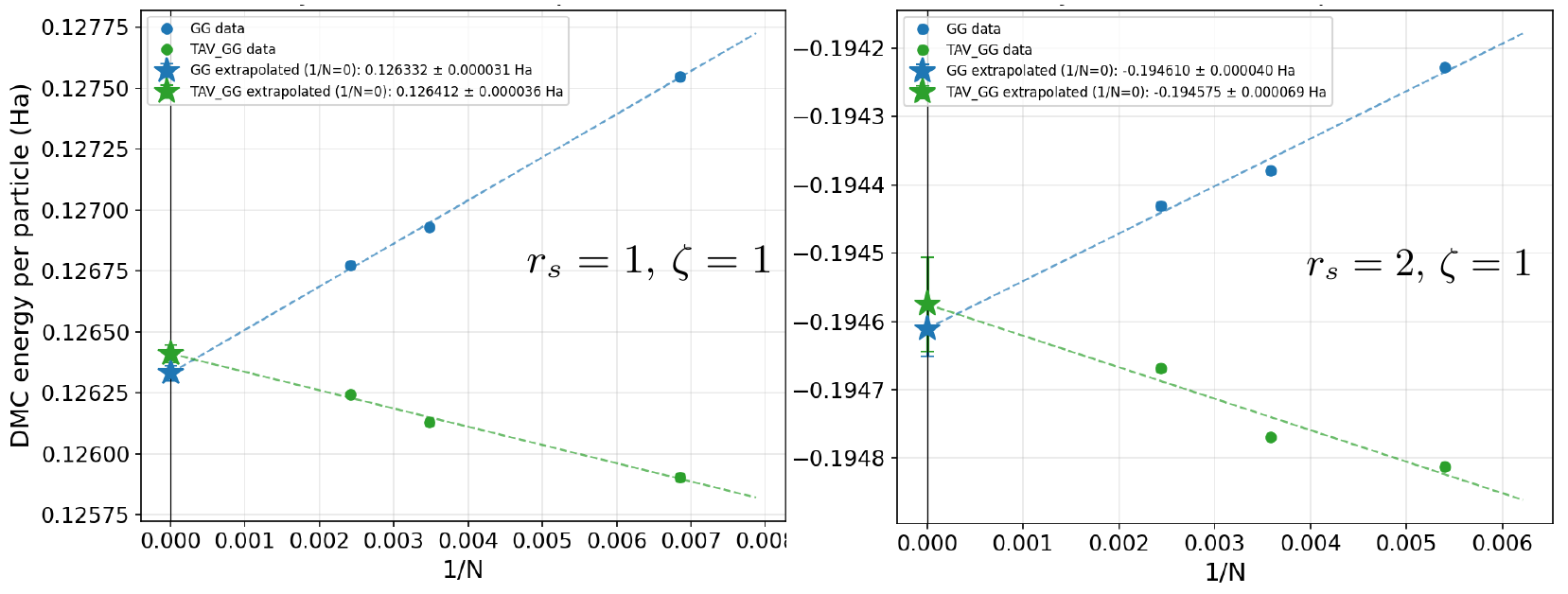}  \\
             \includegraphics[width=0.95\linewidth]{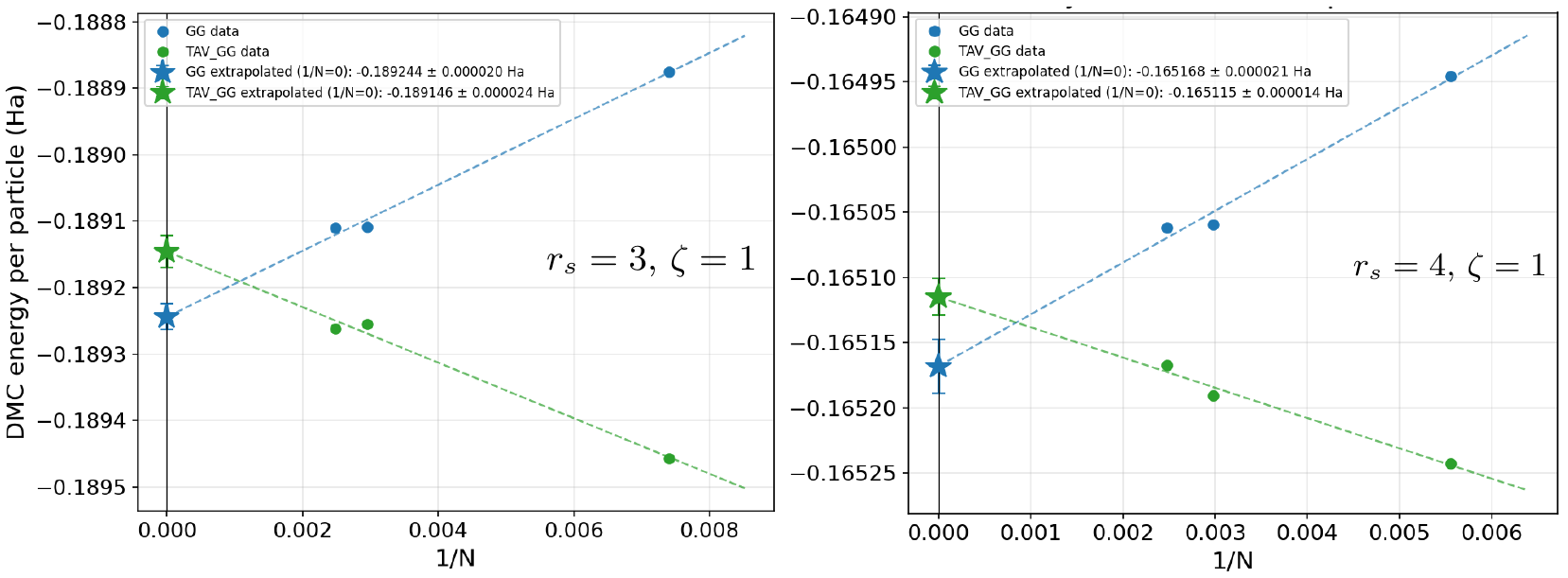}  \\ 
    \end{tabular}
    \caption{Linear extrapolation of GG and TAV-GG DMC energies to $1/N=0$, 2D UEG at $r_s=1,2,3,4$, $\zeta=1$.}
    \label{fig:rs1_extrap_zeta1}
\end{figure*}

We fit $E(N) = E(\infty) + b/N$ to the GG and TAV-GG DMC sequences at each density, at both polarizations. The fits are shown in Figs.~\ref{fig:rs1_extrap} and \ref{fig:rs1_extrap_zeta1}, and Table~\ref{tab:tl} reports the extrapolated $N\to\infty$ energies for both sequences; the quoted uncertainties are the standard errors of the fitted intercepts.

\begin{table}[h]
\centering
\begin{tabular}{c c c c c}
\hline\hline
$\zeta$ & $r_s$ & $E_\infty^{\rm GG}$ (Ha) & $E_\infty^{\rm TAV\text{-}GG}$ (Ha) & $\Delta$ (mHa) \\
\hline
0 & 1 & $-0.209980(888)$ & $-0.210289(438)$ & 0.31 \\
0 & 2 & $-0.258570(109)$ & $-0.258211(112)$ & 0.36 \\
0 & 3 & $-0.212207(78)$  & $-0.211999(47)$  & 0.21 \\
0 & 4 & $-0.175730(136)$ & $-0.175685(101)$ & 0.05 \\
\hline
1 & 1 & $0.126332(31)$   & $0.126412(36)$   & 0.08 \\
1 & 2 & $-0.194610(40)$  & $-0.194575(69)$  & 0.04 \\
1 & 3 & $-0.189244(20)$  & $-0.189146(24)$  & 0.10 \\
1 & 4 & $-0.165168(21)$  & $-0.165115(14)$  & 0.05 \\
\hline\hline
\end{tabular}
\caption{Thermodynamic-limit DMC energies per electron extrapolated from GG and TAV-GG sequences, all four densities, both polarizations. The two extrapolations agree to $0.04$--$0.4$\,mHa, comparable to or smaller than typical statistical uncertainties in DMC benchmarks of this system, and are, if anything, slightly tighter at $\zeta=1$ than at $\zeta=0$.}
\label{tab:tl}
\end{table}

\emph{Choice of extrapolation form.} The leading many-body finite-size correction to the energy of the 2D UEG is expected to scale as $N^{-5/4}$ rather than $1/N$, because the static structure factor behaves as $S(k)\propto k^{3/2}$ at small $k$ in two dimensions, a consequence of the $\sqrt{k}$ plasmon dispersion \cite{Neil08,Chiesa06,Holzmann16}. The $1/N$ form used above is the conventional empirical choice, and over the range $100<N<500$ spanned by 3--5 points per sequence the two forms are not easily distinguished. To establish that our conclusions do not depend on this choice, we repeated every extrapolation with $E(N)=E(\infty)+b\,N^{-5/4}$ (Table~\ref{tab:tl54}). The GG and TAV-GG intercepts again agree to within a fraction of a millihartree in every case, and the agreement is in fact tighter with $N^{-5/4}$ at six of the eight density/polarization combinations; the one case in which it is looser ($r_s=1$, $\zeta=0$) is the sequence with the largest fit residual. We retain $1/N$ as the primary form in the remainder of the paper for two reasons: it is the form against which the slope analysis of Sec.~\ref{sec:opposite_slope} is most transparently interpreted, and the GG sequences are $\Gamma$-point rather than twist-averaged energies, so they carry a residual one-body term (Sec.~\ref{sec:opposite_slope}) for which no particular power law is expected. The central result, agreement between the GG and TAV-GG thermodynamic limits, is the same under either form.

\begin{table}[h]
\centering
\begin{tabular}{c c c c c}
\hline\hline
$\zeta$ & $r_s$ & $E_\infty^{\rm GG}$ (Ha) & $E_\infty^{\rm TAV\text{-}GG}$ (Ha) & $\Delta$ (mHa) \\
\hline
0 & 1 & $-0.209754(724)$ & $-0.210385(350)$ & 0.63 \\
0 & 2 & $-0.258413(86)$  & $-0.258264(88)$  & 0.15 \\
0 & 3 & $-0.212125(57)$  & $-0.212031(37)$  & 0.09 \\
0 & 4 & $-0.175740(110)$ & $-0.175723(79)$  & 0.02 \\
\hline
1 & 1 & $0.126478(8)$    & $0.126349(38)$   & 0.13 \\
1 & 2 & $-0.194560(27)$  & $-0.194610(61)$  & 0.05 \\
1 & 3 & $-0.189203(14)$  & $-0.189181(19)$  & 0.02 \\
1 & 4 & $-0.165139(15)$  & $-0.165132(13)$  & 0.01 \\
\hline\hline
\end{tabular}
\caption{As Table~\ref{tab:tl}, but using the extrapolation form $E(N)=E(\infty)+b\,N^{-5/4}$ appropriate to the leading many-body finite-size correction in two dimensions \cite{Neil08}. The GG and TAV-GG extrapolations agree to $0.01$--$0.6$\,mHa; the conclusions drawn from Table~\ref{tab:tl} are unchanged.}
\label{tab:tl54}
\end{table}

The full VMC, DMC, and HF data at the GG particle numbers underlying the extrapolations above, for every density and both polarizations, are provided in the Supplemental Material \cite{SM}.

\subsection{Testing a one-body correction to the DMC extrapolation}
\label{sec:hf_correction_test}

Since Sec.~\ref{sec:cancellation} establishes that $\Delta E^{\rm HF}_\Gamma(N)$ at Good-$\Gamma$ numbers is a one-body cancellation, we tested directly whether adding this quantity to the DMC energy improves the $1/N$ extrapolation, using the same GG-$N$ sequences as Table~\ref{tab:tl}.  We considered $E_{\rm DMC}(N)+\Delta E^{\rm HF}_\Gamma(N)$ (signed) and $E_{\rm DMC}(N)+|\Delta E^{\rm HF}_\Gamma(N)|$ (absolute value), each fit linearly in $1/N$ as in Sec.~\ref{sec:extrapolation} (Table~\ref{tab:hf_correction_test}).

\begin{table}[h]
\centering
\begin{tabular}{c l c c}
\hline\hline
$r_s$ & quantity & RMS resid.\ (mHa) & intercept (Ha) \\
\hline
1 & $E_{\rm DMC}$                         & 0.277 & $-0.209980(889)$ \\
1 & $E_{\rm DMC}+\Delta E^{\rm HF}_\Gamma$ & 0.288 & $-0.208997(925)$ \\
1 & $E_{\rm DMC}+|\Delta E^{\rm HF}_\Gamma|$ & 0.197 & $-0.210128(632)$ \\
\hline
2 & $E_{\rm DMC}$                         & 0.059 & $-0.258570(109)$ \\
2 & $E_{\rm DMC}+\Delta E^{\rm HF}_\Gamma$ & 0.103 & $-0.258640(190)$ \\
2 & $E_{\rm DMC}+|\Delta E^{\rm HF}_\Gamma|$ & 0.059 & $-0.258549(110)$ \\
\hline
3 & $E_{\rm DMC}$                         & 0.040 & $-0.212206(78)$ \\
3 & $E_{\rm DMC}+\Delta E^{\rm HF}_\Gamma$ & 0.064 & $-0.212244(126)$ \\
3 & $E_{\rm DMC}+|\Delta E^{\rm HF}_\Gamma|$ & 0.047 & $-0.212241(92)$ \\
\hline\hline
\end{tabular}
\caption{RMS residual of a linear-in-$1/N$ fit, and the resulting extrapolated intercept, for the raw DMC energy and for two variants corrected by the $\Gamma$-point HF finite-size error at the same GG-$N$ sequences as Table~\ref{tab:tl}, $\zeta=0$. Neither correction gives a reliable, density-independent improvement (see text).}
\label{tab:hf_correction_test}
\end{table}

The signed correction worsens linearity at every density (the RMS residual increases by $4\%$, $74\%$, and $60\%$ at $r_s=1,2,3$). The absolute-value correction improves linearity at $r_s=1$ ($1.4\times$ smaller residual, with the intercept moving closer to the TAV-GG reference of Table~\ref{tab:tl}, $0.31\to0.16$\,mHa), but leaves it unchanged at $r_s=2$ and slightly worsens it at $r_s=3$. In no case does either correction shift the extrapolated intercept by more than its own fitted uncertainty.

This negative result is itself consistent with the central finding of Secs.~\ref{sec:cancellation}--\ref{sec:hf_structure}, as $\Delta E^{\rm HF}_\Gamma(N)$ at Good-$\Gamma$ numbers is small by cancellation and oscillates in sign with $N$, even within a single density's own GG-$N$ set. It is therefore not a smooth, physically meaningful correction, and there is no reason to expect it to track, or usefully correct for, the DMC energy's own finite-size error. Our results confirm that it does not do so reliably.

\section{Discussion and scope}
\label{sec:discussion}

\emph{Scope.} All results in this work concern the high-density 2D UEG, $r_s=1$--$4$, in its Fermi-liquid regime. We do not claim, and do not test, applicability to real materials, where the number of electrons in a simulation cell is fixed by the lattice and unit-cell choice and the Fermi surface need not be isotropic. Nor do we claim, or test, applicability to the three-dimensional UEG, to the low-density 2D UEG, or to any other model system. The empirical characterization of Sec.~\ref{sec:cancellation} was obtained specifically for the high-density 2D case, and the quantitative behavior established in Sec.~\ref{sec:qmc} should not be assumed to carry over. Extending the GG concept to three dimensions, to lower densities, to real materials, and to anisotropic or multi-sheet Fermi surfaces are each left to separate, future studies. Every statement below is to be read within this scope.

\emph{What the GG approach offers within that scope.} For the 2D UEG, the GG approach has three practical properties relative to twist averaging. First, the $\Gamma$ point is a valid, high-symmetry reciprocal-lattice point for any simulation-cell shape, so identifying GG numbers requires no lattice-specific construction analogous to Baldereschi's mean-value point \cite{Balder73} or the special $k$-point methods of Rajagopal \emph{et al.}\ \cite{Raja94,Raja95}; the same procedure, scan the HF energy over $N$ at $\Gamma$, applies unchanged regardless of cell geometry. Second, the Bloch orbitals at $\Gamma$ can always be chosen real, whereas orbitals at a generic twist are complex; working with real rather than complex orbitals reduces both the memory footprint and the arithmetic cost of every VMC and DMC step. Third, the GG-based sequences used in Sec.~\ref{sec:extrapolation} require only 3--5 independent, real-valued $\Gamma$-point DMC calculations per density, whereas a TA sequence spanning the same range of $N$ requires 50--100 independent, generally complex, twist calculations at each of those $N$ (Sec.~\ref{sec:methods}). A GG-based extrapolation therefore requires far fewer independent calculations and far less workflow overhead than a TA one covering the same $N$ range. Whether this translates into a proportionate saving in total CPU time depends on how sampling effort is distributed among twists, as discussed next, and we make no claim that it does; we also emphasize again that GG is not more accurate than TA at any single fixed $N$.

\emph{Computational cost of twist averaging.} For a fixed target statistical uncertainty, Monte Carlo sampling effort can in principle be distributed among twists, so the cost of twist averaging does not scale linearly with the number of twists, and the total number of Monte Carlo steps needed for a TA energy of given precision need not greatly exceed that for a single $\Gamma$-point energy of the same precision. Twist averaging nonetheless requires separate equilibration, walker populations, orbital sets, and data handling for each twist, and may require twist-dependent wave-function preparation or optimization; it therefore carries non-negligible setup and workflow overhead relative to a single, carefully selected $\Gamma$-point calculation. This overhead is system- and implementation-dependent, and is relatively modest for the UEG studied here. The honest summary is that GG reduces the number and complexity of calculations, not necessarily the number of Monte Carlo samples.

\emph{Relation to prior special-point methods.} The GG criterion explored here is conceptually related to, but distinct from, the exact special twist (EST) method of Dagrada \emph{et al.}\ \cite{Dagrada16} and the mean-field-based correction of Kwee, Zhang, and Krakauer \cite{Kwee08}. EST continuously tunes a single twist angle at fixed $N$ to drive the mean-field FS error to an arbitrarily small, controlled value; it is, by construction, always available and always accurate to the desired tolerance. GG, by contrast, fixes the twist at $\Gamma$ and searches over the discrete set of accessible $N$, achieving a small one-body error only at specific, sparsely distributed particle numbers, and, as shown in Sec.~\ref{sec:hf}, by cancellation rather than by directly minimizing the error. We do not claim GG is more accurate or more broadly applicable than EST or than twist averaging. The shell oscillation of the $\Gamma$-point energy (Fig.~\ref{fig:HF}) has been known since the introduction of twist averaging \cite{Lin01}; what we add is its decomposition into kinetic and exchange components at the particle numbers where it nearly vanishes, the demonstration that this vanishing is a cancellation rather than a convergence, and a direct test, against twist-averaged DMC at eight density/polarization combinations, of what such particle numbers can and cannot deliver at the correlated level (Sec.~\ref{sec:qmc}).

\section{Conclusion}
We have characterized Good-$\Gamma$ (GG) particle numbers -- system sizes at which the $\Gamma$-point HF energy of the high-density 2D UEG lies anomalously close to its thermodynamic-limit value -- and shown, at $r_s=1,2,3,4$ and at both unpolarized ($\zeta=0$) and fully spin-polarized ($\zeta=1$) polarization, that this near-cancellation involves individually large kinetic- and exchange-energy errors that cancel in the total. The ratio between them is negative and grows with $r_s$ in close to direct proportion at every density and polarization studied, though its overall scale is somewhat larger at $\zeta=0$ than at $\zeta=1$ at the same $r_s$. At all four densities and both polarizations, we additionally tested the correlated DMC level and found that a GG-based sequence, once selected with $N$ reasonably spread in $1/N$, extrapolates to the thermodynamic limit in good agreement with twist-averaged DMC extrapolation over the same range, whether a $1/N$ or an $N^{-5/4}$ extrapolation form is used. We clarify the reason behind the opposite-sign finite-size trends of GG and TA sequences. Within the scope of this work -- the high-density 2D UEG -- a well-spaced GG sequence offers a practical route to the thermodynamic limit that requires no lattice-specific construction, always admits a real wave function, and needs far fewer independent DMC calculations than a TA sequence covering the same $N$ range.

\section{Acknowledgment}
S.A, A.P, and M.S.B acknowledge the support of the Leverhulme Trust under the grant agreement RPG-2023-253. S. Azadi and T.D. K\"{u}hne acknowledge the computing time provided to them on the high-performance computers Noctua2 at the NHR Center in Paderborn (PC2). The data that support the findings of this article are openly available \cite{github}, embargo periods may apply.

\bibliography{main}
\end{document}